\documentclass{dcds}
  \usepackage{paralist}
  \usepackage{graphics} 
  \usepackage{epsfig} 
 \usepackage{graphicx}
\usepackage{epsf,epstopdf,psfrag}
\usepackage{amsmath,amscd,amssymb,dsfont}
\usepackage[colorlinks=true]{hyperref}
\hypersetup{urlcolor=blue, citecolor=red}

\def\currentvolume{28}

    \def\ppages{1--XX}
     \def\DOI{10.3934/dcds.2010.28.xx}

\theoremstyle{definition}

\title[Mixing Measures and Effective Diffusion]
      {Models \& Measures of Mixing \& Effective Diffusion}

\author[Zhi Lin, Katar{\'\i}na Bo{\v d}ov{\' a} and Charles R. Doering]{}

\subjclass{76F25, 76M50, 74Q20, 76R05.}
 \keywords{Transport \& mixing; Stirring \& mixing; Turbulent diffusion.}

 \email{linxa003@ima.umn.edu}
 \email{Katarina.Bodova@fmph.uniba.sk}
 \email{doering@umich.edu}

\thanks{This work was supported in part by NSF awards DMS--0553487, PHY--555324, and PHY--0855335.}

\begin{document}
\maketitle

\centerline{\scshape Zhi Lin }
\medskip
{\footnotesize
 \centerline{Institute for Mathematics \& Its Applications}
 \centerline{University of Minnesota, Minneapolis, MN 55455, USA}
} 

\bigskip

\centerline{\scshape Katar{\'\i}na Bo{\v d}ov{\'a}}
\medskip
{\footnotesize
 \centerline{Department of Applied Mathematics \& Statistics}
 \centerline{Faculty of Mathematics, Physics and Informatics}
 \centerline{Comenius University, 84248 Bratislava, Slovakia}
}

\bigskip
\centerline{\scshape Charles R. Doering}
\medskip
{\footnotesize
 \centerline{Department of Mathematics, University of Michigan}
  \centerline{Ann Arbor, MI 48109-1043, USA}
 \centerline{and}
 \centerline{Department of Physics and Michigan Center for Theoretical Physics}
  \centerline{University of Michigan, Ann Arbor, MI 48109-1040, USA}
 \centerline{and}
 \centerline{Center for the Study of Complex System, University of Michigan}
   \centerline{Ann Arbor, MI 48109-1107, USA}
}

\begin{abstract}
Mixing a passive scalar field by stirring can be measured in a variety of ways including tracer particle dispersion, via the flux-gradient relationship, or by suppression of scalar concentration variations in the presence of inhomogeneous sources and sinks.
The mixing efficiency or efficacy of a particular flow is often expressed in terms of enhanced diffusivity and quantified as an effective diffusion coefficient.
In this work we compare and contrast several notions of effective diffusivity.
We thoroughly examine the fundamental case of a steady sinusoidal shear flow mixing a scalar sustained by a steady sinusoidal source-sink distribution to explore apparent quantitative inconsistencies among the measures.
Ultimately the conflicts are attributed to the noncommutative asymptotic limits of large P$\acute{\text{e}}$clet number and large length-scale separation.
We then propose another approach, a generalization of Batchelor's 1949 theory of diffusion in homogeneous turbulence, that helps unify the particle dispersion and concentration variance suppression measures.
\end{abstract}

\section{Introduction}
Flow-enhanced mixing is an important phenomenon in natural systems varying in size from as small as human cells to as large as the atmosphere and the ocean and beyond \cite{papanicolaou,gollub,taylor}.
The enhancement of molecular mixing by stirring can be observed even for simple laminar flows, and a quantitative understanding of fundamental mechanisms and properties of mixing processes is key to accurate modeling of these systems.

Passive scalars are mathematically idealized entities that serve as proxies to formulate and investigate this problem.
Given its initial location, the trajectory of a passive tracer in $\mathds{R}^d$ advected by a prescribed flow field $\vec{u}(\vec{x},t)$ is described by the stochastic differential equation
\begin{equation}
d\vec{X}(t) = \vec{u}(\vec{X}(t),t) dt + \sqrt{2\kappa}\; d\vec{W}(t),
\label{masterSDE}
\end{equation}
where $\kappa$ is the molecular diffusivity and $\vec{W}$ is canonical $d$-dimensional Brownian motion.  Equivalently, the Fokker-Planck equation that governs the evolution of the scalar concentration field $T(\vec{x},t)$ stirred by an incompressible ($\nabla\cdot\vec{u}=0$) flow is the advection-diffusion equation
\begin{equation}
\partial_t T + \vec{u} \cdot \nabla T = \kappa \Delta T \,
\label{mastereq}
\end{equation}
supplemented with initial concentration $T(\vec{x},0)$ and appropriate boundary conditions.  In many applications, the scalar field is constantly replenished and depleted by sources and sinks.  Consequently, the homogeneous partial differential equation (\ref{mastereq}) would be appended with an inhomogeneous term corresponding to the source-sink distribution whose relationship with the stirring further adds to the mathematical complexity of the problem.

In this work we first review several different measures established in the literature to characterize the effects of stirring on enhanced mixing.  Each measure is associated with a specific physical setting and is amenable to different mathematical analysis and/or approximation.  Approximations adopting distinct noncommutative limiting procedures for control parameters may yield conflicting predictions.  We then thoroughly investigate a fundamental example of this problem, that of a simple steady shear flow stirring a scalar sustained by a monochromatic source-sink distribution, to expose an apparent contradiction between the mixing measures when the high-P$\acute{\text{e}}$clet and large length-scale-separation limits are exchanged.  Finally, we propose a mathematical framework that utilizes information about single particle dispersion, a generalization of Batchelor's 1949 theory \cite{B49} of diffusion in a field of homogeneous isotropic turbulence, to accurately predict scalar concentration variance reduction by the (in this case, inhomogeneous and anisotropic) flow.  This new approach produces a uniformly valid dependence of the effective diffusion on the control parameters, reconciling the apparent inconsistencies.

\section{Mixing measures}
A conventional modeling approach, and the one we will focus on here, is to describe the flow-enhanced mixing by replacing the advection-diffusion operator with a effective diffusive operator, i.e.,
\begin{equation}
\kappa \Delta T-\vec{u} \cdot \nabla T \ \to \
\nabla\cdot(\mathbf{K}^{\text{eff}}\cdot\nabla T)
\label{effrep}
\end{equation}
where $\mathbf{K}^{\text{eff}}$ is an effective diffusivity tensor designed to capture some specific feature(s) of the mixing process.

One such feature is transient passive particle dispersion with $\mathbf{K}^{\text{eff}}$ defined by
\begin{equation}
K_{ij}^{\text{eff}}=\frac12\lim_{t\to \infty} \frac{d}{dt} \mathds{E}[ (X_i(t)-X_i(0))(X_j(t)-X_j(0))],\quad i,j=1,\cdots,d
\label{defhomokeff}
\end{equation}
where $\vec{X}(t)=(X_1(t),\cdots,X_d(t))^T$ is the trajectory described by (\ref{masterSDE}) and the statistical average $\mathds{E}[\cdot]$ is taken over all Brownian paths $\vec{W}(t)$.
The effective diffusion tensor $K_{ij}$ is often defined as $\lim_{t\to \infty} \frac{1}{2t} \mathds{E}[ (X_i(t)-X_i(0))(X_j(t)-X_j(0))]$, but these are equivalent when the covariance elements converge grow no more than linearly with time.

The steady sinusoidal shear $\vec{u}=\hat{i}\sqrt{2}\,U\sin(k_uy)$, where $\hat{i}$ is the unit vector in the $x-$direction, is a case in point.
The effective diffusivity tensor defined by the long-time dispersive behavior of particles is the Taylor dispersion result \cite{taylor53}
\begin{equation}\mathbf{K}^{\text{eff}}=\begin{pmatrix}\kappa (1+\textit{Pe}^2)&0\\0&\kappa\end{pmatrix}
\label{homolimit}
\end{equation}
where the strength of the advection is gauged by the non-dimensional P$\acute{\text{e}}$clet number $\textit{Pe}=Ul_u/\kappa$ with the flow-characteristic length-scale $l_u=k_u^{-1}$.

In the absence of steady sources and sinks, and when the separation between the characteristic length-scale of the scalar density distribution, call it $l_d$, and that of the flow, $l_u$, is very large, homogenization theory \cite{papanicolaou,majdaMc, majda_kramer} asserts that the scalar concentration evolves according to a diffusion equation with effective diffusion tensor $\mathbf{K}^{\text{eff}}$ for a broad collection of deterministic and stochastic flows.
The normalized tensor $\kappa^{-1} \mathbf{K}^{\text{eff}}$ emerges naturally as a dimensionless measure of the mixing efficacy of the stirring, and previous analysis for the $l_d/l_u\to\infty$ homogenization limit has shown that each component in the normalized tensor is bounded above in terms of the P$\acute{\text{e}}$clet number according to
\begin{equation}E_{ij}^{\text{HT}}:=
\frac{K^{\text{eff}}_{ij}}{\kappa}\le 1+\textit{Pe}^2.
\label{boundht}
\end{equation}
Taylor dispersion, and in particular the steady sinusoidal shear flow, saturates the bound for $K^{\text{eff}}_{11}$.

Homogenization theory explicitly assumes the large length-scale separation between the flow and the scalar field.
In one incarnation a steady large-scale gradient is imposed in one spatial direction to formulate the so-called ``cell problem" \cite{gollub}.
This is implemented theoretically by writing $T=-Gx+\theta$ so the advection-diffusion equation (\ref{mastereq}) becomes the evolution equation for the concentration fluctuation field $\theta$:
\begin{equation}\partial_t\theta+ \vec{u} \cdot \nabla \theta = \kappa \Delta \theta+G\,(\hat{i}\cdot\vec{u}).
\label{mastereqfg}
\end{equation}
The flow and fluctuation fields are then assumed to be periodic and mean-zero on a cell of size $l_u$ across which the ``mean" scalar gradient $G$ is held constant.
This model serves as the starting point for many studies of turbulent mixing \cite{joergsreeni}.

The solution of (\ref{mastereqfg}) provides another, potentially distinct, measure of mixing enhancement in terms of the scalar flux-gradient relationship.
If the effective diffusion coefficient $K^{\text{eff}'}_{11}$ is defined as the ratio of the mean scalar flux in the $x-$direction to the mean scalar gradient in the $x-$direction, then
\begin{equation}
E_{11}^{\text{FG}}:= \frac{K^{\text{eff}'}_{11}}{\kappa} =\frac{\langle \hat{i}\cdot ( \vec{u}\, T -\kappa \nabla T)\rangle}{\kappa G} = 1+\frac{\langle|\nabla\theta|^2\rangle}{G^2}\le1+\textit{Pe}^2
\label{boundfg}
\end{equation}where $\langle\cdot\rangle$ denotes the long time and spatial average within a periodic cell.
The second expression for $E_{11}^{\text{FG}}$ in terms of $\theta$ follows from the time-cell average of $\theta$ times (\ref{mastereqfg}) which implies that $\kappa \langle|\nabla\theta|^2\rangle=G\langle \theta (\hat{i}\cdot\vec{u}) \rangle$, and the upper bound follows from
\begin{eqnarray}
\frac{\langle|\nabla\theta|^2\rangle}{G}=\frac{\langle \theta (\hat{i}\cdot\vec{u})\rangle}{\kappa}
&=&-\frac{\langle \nabla \theta \cdot \nabla^{-1} (\hat{i}\cdot\vec{u})\rangle}{\kappa}\nonumber \\
&\le& \frac{\langle|\nabla\theta|^2\rangle^{\frac{1}{2}}\langle|\nabla^{-1}\vec{u}|^2\rangle^{\frac{1}{2}}}{\kappa}
=\textit{Pe}\,\langle|\nabla\theta|^2\rangle^{\frac{1}{2}}
\end{eqnarray}
where the P\'eclet number is defined in terms of the rms velocity $U$ and the characteristic length-scale of the flow $l_u$ as
\begin{equation}U^2:=\langle |\vec{u}|^2\rangle, \quad  l_u^2:={\langle |\nabla^{-1}\vec{u}|^2\rangle}/{\langle |\vec{u}|^2\rangle}\,.
\label{defUlu}
\end{equation}Here the inverse gradient operator $\nabla^{-1}$ has the Fourier symbol $-i|\vec{k}|^{-2}\vec{k}$ operating on mean-zero functions.  The same shear flow $\vec{u}=\hat{i}\sqrt{2}\,U\sin(k_uy)$ that saturates the bound in (\ref{boundht}) also saturates the bound in (\ref{boundfg}) because the steady-state solution to (\ref{mastereqfg}) is $\theta(y)=\sqrt{2}\,GU\sin(k_uy)/\kappa k_u^2$.
In homogenization theory, the effective diffusivity defined in (\ref{defhomokeff}) is often computed by solving (\ref{mastereqfg}).

Another mixing measure was recently introduced motivated by problems with spatially inhomogeneous scalar sources and sinks \cite{doering,plasting,trouble,TDG}.
Transporting particles from sources to sinks by advection may help to suppress the inevitable inhomogeneities in the scalar field beyond that which molecular diffusion can accomplish alone.
Inhomogeneities in the scalar concentration may be measured by its space-time averaged variance, and stirring efficacies can be defined in terms of the suppression of scalar variance on various spatial length-scales.
In particular, in the presence of a given source-sink distribution $s(\vec{x},t)$, Thiffeault \emph{et al.} introduced the notion of scale dependent \emph{equivalent diffusivities} \cite{doering,TDG}
\begin{equation}
\kappa^{\text{eq}}_p \ = \ \sqrt{\frac{\langle|\nabla^p\Delta^{-1}s|^2\rangle}{\langle|\nabla^p\theta|^2\rangle}} \ = \
\kappa\,\sqrt{\frac{\langle|\nabla^p\theta_0|^2\rangle}{\langle|\nabla^p\theta|^2\rangle}}
\label{defkeq}
\end{equation}
where the unstirred ``reference" scalar field $\theta_0$ satisfies
\begin{equation}
\partial_t\theta_0=\kappa\Delta\theta_0+s
\end{equation}
while the stirred scalar field $\theta$ satisfies
\begin{equation}\partial_t\theta+\vec{u}\cdot\nabla\theta=\kappa\Delta\theta+s.
\label{eqtheta}
\end{equation}
The associated mixing efficacies
\begin{equation}
\mathcal{E}_p:=\frac{\kappa^{\text{eq}}_p}{\kappa}
\label{defmixeff}
\end{equation}
are measures of the statistical steady-state flow-enhanced concentration variance reduction field on small ($p=1$), intermediate ($p=0$), and large ($p=-1$) spatial scales.
Here the source-sink distribution $s(\vec{x},t)$ is, without loss of generality, spatially mean 0.
This notion parametrizes the flow-enhanced mixing by the effective or ``equivalent'' molecular diffusion coefficient that achieves the same level of scalar concentration variance suppression that the stirring supplies.

Rigorous analysis shows that given a stationary source-sink distribution $s(\vec{x})$, the efficacies and equivalent diffusivities are bounded by
\begin{equation}
\mathcal{E}_p \le \widetilde{C}\,\textit{Pe}
\label{boundvarred}
\end{equation}
as $\textit{Pe} \rightarrow \infty$ for a wide class of deterministic and stochastic flows with a prefactor $\widetilde{C}$ depending on $l_s/l_u$, where $l_s$ is a length-scale characterizing the source-sink distribution.  This linear-in-$\textit{Pe}$ bound can be saturated for some source-sink and flow combinations, as shown by Plasting and Young \cite{plasting}.  Moreover, for time-dependent flows with statistical homogeneity and isotropy properties often associated with fully developed turbulence, even smaller estimates for $\mathcal{E}_p$, i.e., asymptotic upper bounds $\lesssim \textit{Pe}^\alpha$ with $\alpha<1$, hold for certain classes of source-sink distributions and depend on the spatial dimension \cite{doering}.
These estimates have also been shown to be sharp \cite{okabe}.

The discrepancy between the $\textit{Pe}^2$ scaling (\ref{boundht}) and the $\textit{Pe}^1$ scaling (\ref{boundvarred}) of the diffusion enhancement factors casts doubt on the universal applicability of the approaches and their associated mixing measures, and raises questions about modeling mixing by (\ref{mastereqfg}) when $\textit{Pe}$ is large.
This discrepancy may to some extent be attributed to the necessity of a scale separation between the tracer concentration and flow fields in the basic particle dispersion analysis, and its inevitable implication in flux-gradient models.
Equivalently, maximally enhanced diffusion in these approaches requires time to develop, time which may be as large as $\mathcal{O}(l_u^2/\kappa)$, the effective ``mean free time'' for a typical tracer from the initial distribution to move by molecular diffusion onto a streamline in another direction.
Said differently, Taylor dispersion may require a long time to emerge \cite{youngjones}.

In the presence of sustained scalar sources and sinks this separation of time scales may never effectively be achieved: a relevant time scale in the sourced problem is the lesser of $l_u^2/\kappa$, the time for enhanced Taylor-like dispersion to appear, and $l_s/U$, the time it takes for a particle to be advected by the flow from a source to a sink.
If $l_s/U < l_u^2/\kappa$, the bulk of the scalar variance may be dominated by particles that are most recently replenished and depleted rather than by particles that have been in the system for a long time, and are thus relatively well mixed.  Transient features of particle dispersion cannot be neglected and the simple parameterization of the advection-diffusion operator with a stationary tensor may not be valid.
As a result, mixing efficacies for source-sink problems may differ from those deduced from transient mixing problems when $l_s/U < l_u^2/\kappa$, i.e., when at high P\'eclet numbers when $Pe > l_s/l_u$.
The specific example analyzed in detail in the next section precisely illustrates this noncommutativity of the large length-scale-separation limit and the large P$\acute{\text{e}}$clet number limit.
These two distinct asymptotic limits can produce different predictions for the effective diffusion scaling.

\section{Explicit example:  Sinusoidal source-sink and shearing}

\begin{figure}
\centering
     \centering
     \hskip -1.5cm
     \includegraphics[width=11cm,height=9cm]{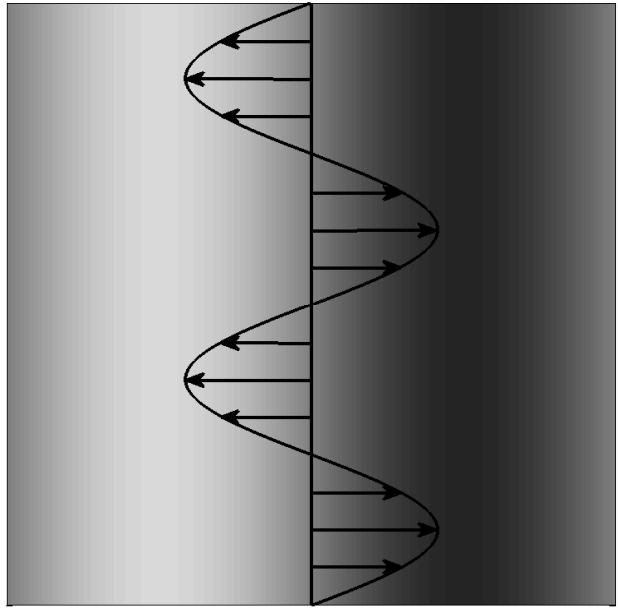}
     \caption{\label{sinflow}The sinusoidal shear $\vec{u}=\hat{i}\sqrt{2}\,U\sin(k_uy)$ with source distribution $s(\vec{x},t)=\sqrt{2}\,S\sin(k_s x)$ in the shaded background.}
\end{figure}

Consider the simplest nontrivial case where the steady single length-scale flow
\begin{equation}
\vec{u}(\vec{x},t)=\hat{i}\sqrt{2}\,U\sin(k_u y)
\end{equation}
is stirring the steady single-scale source
\begin{equation}
s(\vec{x},t)=\sqrt{2}\,S\sin(k_s x)
\end{equation}
as depicted in Figure \ref{sinflow}.  The non-dimensional control parameters are the P$\acute{\text{e}}$clet number $\textit{Pe}$ and the length-scale separation ratio $r$, which we define as
\begin{equation}\textit{Pe}=\frac{U}{\kappa k_u},\quad \quad r=\frac{k_u}{k_s}=\frac{l_s}{l_u}\,.
\label{defPeR}
\end{equation}The reference, unstirred, steady scalar concentration is $\theta_0(x)=\frac{\sqrt{2}\,S}{\kappa k_s^2}\,\sin(k_s x)$ while the exact steady ($t \rightarrow \infty$) stirred solution $\theta_\infty(x,y)$ solving
\begin{equation}\sqrt{2}\,U\sin(k_uy)\partial_x{\theta_\infty}=\kappa\Delta \theta_\infty+\sqrt{2}\,S\sin(k_s x)
\label{eqthetainf}
\end{equation}has the form
\begin{equation}\theta_\infty(x,y)=f(y)\sin(k_sx)+g(y)\cos(k_sx)
\label{thetainfdecomp}
\end{equation}where $f$ and $g$ are $\frac{2\pi}{k_u}-$periodic functions that can be computed via numerical or asymptotic methods.

\bigskip

\begin{figure}
     \centering
     \includegraphics[width=7.8cm,height=9.672cm,angle=-90]{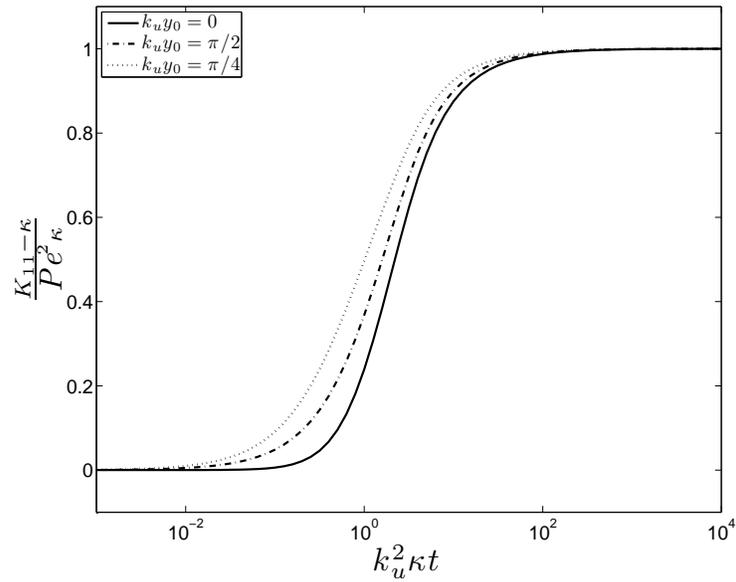}
     \caption{\label{k11}$\frac{K_{11}-\kappa }{\textrm{Pe}^2\kappa}$ vs $k_u^2\kappa t$ for $k_uy_0=0,\frac{\pi}{2},\frac{\pi}{4}.$}
\end{figure}

Different dependencies of the different mixing enhancement measures at large P$\acute{\text{e}}$clet numbers are by now well documented  \cite{majda_kramer,trouble,taylor53,youngjones}.
In the limit $r\to\infty$ the direct approximate solution obtained by homogenization theory, i.e., the solution of the steady inhomogeneous diffusion equations effective diffusivity given by the particle dispersion or flux-gradient definition, is simply
\begin{equation}\theta_{\text{HT}}(x)=\frac{\sqrt{2}\; S\; \sin(k_s x)}{\kappa k_s^2(1+\textit{Pe}^2)}
=\frac{\theta_0(x)}{1+\textit{Pe}^2}
.
\label{htsol}
\end{equation}
This suggests that the homogenization theory approximation is
\begin{equation}\mathcal{E}_p^{\text{HT}}=\sqrt{\frac{\langle|\nabla^p\theta_0|^2\rangle}{\langle|\nabla^p\theta_{\text{HT}}|^2\rangle}}= 1+\textit{Pe}^2
\label{limitht}
\end{equation}
for all $p$, although careful application of homogenization arguments to the {\it gradient} of the scalar field reduces the homogenization theory prediction for $\mathcal{E}_1$ to $\mathcal{O}(\textit{Pe})$ as $\textit{Pe} \rightarrow \infty$ \cite{KKS}.

To see (\ref{limitht}) we derive the explicit formula for tracer particle position covariance from the solutions of the stochastic differential equations written as
\begin{equation}
\begin{array}{l}X(t)=x_0+\sqrt{2}U\,\int_0^t \sin(k_uY(s))ds+\sqrt{2\kappa}\;W_1(t),\\[1.2ex]
Y(t)=y_0+\sqrt{2\kappa}\;W_2(t)
\label{sde}
\end{array}
\end{equation}
where $\vec{X}(t=0)=(x_0,y_0)^T$ and $W_1$ and $W_2$ are two independent Brownian motions.
The entries in the 2-by-2 covariance tensor $\mathbf{C}(y_0,t)$ are computed as
\begin{eqnarray}
\mathbf{C}_{11}(y_0,t)&=&\mathds{E} \big[\big(X(t)-x_0\big)^2\big] \nonumber \\
&=& \ 2\kappa t + 2U^2\int_{[0,t]^2}\mathds{E} \big[\sin(k_uY(s))\sin(k_uY(\tau))\big] ds\,d\tau \nonumber \\
= \ \ \ 2\kappa t &+& \frac{2U^2}{\kappa^2k_u^4}\Big[k_u^2\kappa t-1+e^{-k_u^2\kappa t}-\frac{3-4e^{-k_u^2\kappa t}+e^{-4k_u^2\kappa t}}{12}\,\cos(2k_uy_0)\Big], \nonumber \\
&\sim& \,2\big(\kappa+\frac{U^2}{\kappa k_u^2}\big)\; t, \quad \text{as}\ \ t\to\infty,
\label{Eentries11}
\end{eqnarray}
\begin{eqnarray}
\mathbf{C}_{12}(y_0,t)&=&\mathds{E} \big[ \big(X(t)-x_0\big)\big(Y(t)-y_0\big)\big]\nonumber \\
&=& 2U\sqrt{\kappa}\int_0^t\mathds{E} \big[W_2(t)\sin(k_uY(s))\big] ds \nonumber \\
&=& \,\frac{2\sqrt{2}\,U\cos(k_uy_0)t}{k_u}\Big(\frac{1-e^{-k_u^2\kappa t}}{k_u^2\kappa t}-e^{-k_u^2\kappa t}\Big)\to 0,\quad t\to\infty,  \label{Eentries12} \\
\nonumber \\
\mathbf{C}_{22}(y_0,t) &=& \mathds{E} \big[ \big(Y(t)-y_0\big)^2\big]
\ \ = \ \ 2\kappa t.
\label{Eentries22}
\end{eqnarray}
The homogenization theory approximation (\ref{homolimit}), i.e, classical Taylor dispersion, follows from the long time limits of these quantities.
Figure \ref{k11} shows the temporal evolution of $K_{11}$ for different values of $y_0$.
Note that it takes time $\approx 100/k_u^2 \kappa$ for the full $\textit{Pe}^2$ enhancement to emerge.

In contrast, it has been shown \cite{TDG} that the efficacy $\mathcal{E}_{0}$ of any flow on the torus, steady or time-dependent, stirring the simple monochromatic source-sink distribution is bounded according to
\begin{equation}\mathcal{E}_{0}\le \sqrt{1+r^2\textit{Pe}^2}.
\label{boundrpe}
\end{equation}
Because $r\textit{Pe} = U/\kappa k_s$, this upper bound does not depend on any length-scales in the flow.
In fact, for this source-sink distribution this bound is saturated by a steady, spatially uniform wind directly blowing source to sink and sink to source.

For the sinusoidal shear flow $\vec{u}(\vec{x},t)=\hat{i}\sqrt{2}\,U\sin(k_u y)$, a detailed high-$Pe$ asymptotic analysis for the exact steady-state solution $\theta_\infty$  \cite{trouble}, what we will refer to here as Internal Layer Theory (ILT), shows
\begin{equation}
\mathcal{E}_{0}\sim r^{7/6}\textit{Pe}^{5/6}, \quad \quad
\mathcal{E}_{-1}\sim r\,\textit{Pe}, \quad \quad
\mathcal{E}_{+1}\sim r^{1/2}\textit{Pe}^{1/2}
\label{asympilt}
\end{equation}
as $\textit{Pe}\to\infty$ at fixed $r$.
Comparing the asymptotic bounds (\ref{limitht}) and (\ref{asympilt}), it is clear that with two control parameters, $\textit{Pe}$ and $r$, the large-$\textit{Pe}$ asymptotics does \emph{not} commute with large-$r$ asymptotics.   As a result, the dependence of $\mathcal{E}_0$ on large P$\acute{\text{e}}$clet numbers has two distinguished regimes that cross over near $r=\textit{Pe}$.

\begin{figure}
  \centering
  \includegraphics[width=8cm,height=7.5cm]{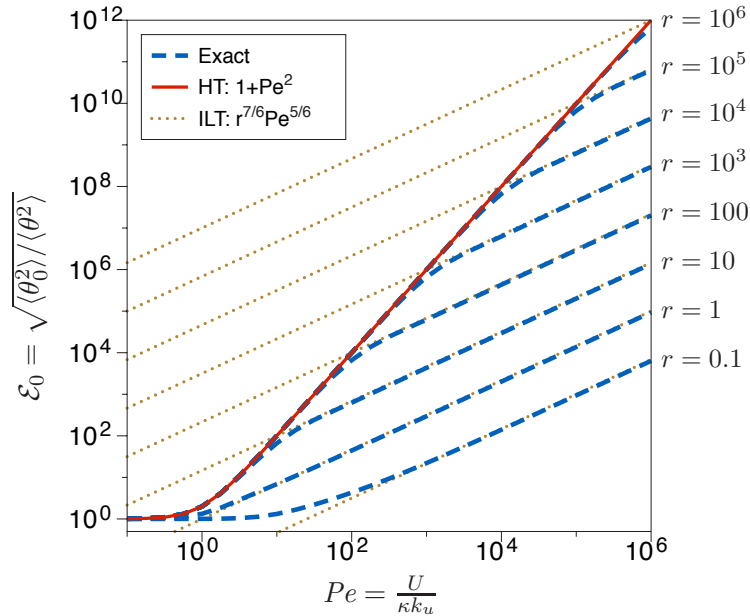}
  \vskip -2.6cm
  \hskip -9cm
  \rotatebox{90}{\makebox[0pt][l]{\large $\mathcal{E}_0=\sqrt{\langle\theta_0^2\rangle/\langle\theta^2\rangle}$}}
  \vskip -5.4cm
  \hskip 7.8cm
  \makebox[0pt][l]{\normalsize $r=10^6$}
  \vskip 0.2cm
  \hskip 7.8cm
  \makebox[0pt][l]{\normalsize $r=10^5$}
  \vskip 0.2cm
  \hskip 7.8cm
  \makebox[0pt][l]{\normalsize $r=10^4$}
  \vskip 0.25cm
  \hskip 7.8cm
  \makebox[0pt][l]{\normalsize $r=10^3$}
  \vskip 0.23cm
  \hskip 7.8cm
  \makebox[0pt][l]{\normalsize $r=100$}
  \vskip 0.23cm
  \hskip 7.8cm
  \makebox[0pt][l]{\normalsize $r=10$}
  \vskip 0.23cm
  \hskip 7.8cm
  \makebox[0pt][l]{\normalsize $r=1$}
  \vskip 0.23cm
  \hskip 7.8cm
  \makebox[0pt][l]{\normalsize $r=0.1$}
  \vskip 2.7cm
  \hskip -1.2cm
  \makebox[0pt][l]{\large $\textit{Pe}=\frac{U}{\kappa k_u}$}
  \caption{\label{e0plot}$\mathcal{E}_0$ vs $\textit{Pe}$ for $r=10^m,\;m=-1,0,\cdots,6.$}
\end{figure}

Figure \ref{e0plot} illustrates the accuracy of different theoretical estimates for $\mathcal{E}_0$ by comparison with the exact value for $10^{-1}\le\textit{Pe}\le10^6$ and $r=10^m$, $m=-1,\cdots,6$.  The exact behavior of $\mathcal{E}_0$ shows that the two limits, $\textit{Pe}\to\infty$ and $r\to\infty$, do not generally commute and thus lead to two distinct asymptotic regimes of the $\textit{Pe}$ dependence: the $1+\textit{Pe}^2$ regime for $\textit{Pe} \lesssim r$ and the $r^{7/6}\textit{Pe}^{5/6}$ regime for $\textit{Pe} \gtrsim r$.  The question we now turn to is whether transient particle dispersion information can be utilized to correctly predict steady-state concentration variance suppression in the presence of steady sources and sinks.

\section{Dispersion-diffusion theory}
In order to reconcile the notions of effective diffusion in terms of particle dispersion, on the one hand, and source-sink sustained scalar variance suppression, on the other hand, we propose what we call Dispersion-Diffusion Theory (DDT).
Specifically, we retain the dependence of the effective diffusivity tensor on time \cite{B49} and initial location \cite{youngjones} by defining
\begin{equation}
\mathbf{K}(\vec{x}_0,t):=\frac12 \frac{d}{dt}\mathbf{C}(\vec{x}_0,t) = \frac12 \frac{d}{dt} \mathds{E}[ (X_i(t)-X_i(0))(X_j(t)-X_j(0)) ].
\label{defK}
\end{equation}
Then without sources and sinks, it is hypothesized that the probability density of a single passive particle may be approximated by the diffusion equation
\begin{equation}
\frac{\partial}{\partial t} \rho(\vec{x},t ; \vec{x}_0,t_0) = \frac{\partial}{\partial x_i} K_{ij}(\vec{x}_0,t-t_0)  \frac{\partial}{\partial x_j} \rho(\vec{x},t ; \vec{x}_0,t_0)
\end{equation}
with initial distribution $\rho(x,t_0 ; \vec{x}_0,t_0) = \delta(\vec{x}-\vec{x}_0)$.  For spatially periodic problems with period $L$, this fundamental solution is
\begin{equation}
\rho(\vec{x},t ; \vec{x}_0,t_0) = \frac1{L^d} \sum_{\frac{L\vec{k}}{2\pi}\in\, \mathds{Z}^d} e^{i\vec{k}\cdot(\vec{x}-\vec{x}_0)-\frac12\,\vec{k} \cdot \mathbf{C}(\vec{x}_0,t-t_0) \cdot \vec{k}}.
\label{rhosol}
\end{equation}
Now we propose to approximate the solution to (\ref{eqtheta}) with a source-sink distribution $s(\vec{x},t)$ by the integral
\begin{equation}
\theta_{\text{DDT}}(\vec{x},t) = \int_{0}^t dt_0 \int_{[0,L]^d} d \vec{x}_0 \rho(\vec{x},t ; \vec{x}_0,t_0) s(\vec{x}_0,t_0)\; .
\label{ansatz}
\end{equation}
That is, we simply apply the principle of linear superposition to the particle density introduced (or depleted) at all positions $\vec{x}_0$ at all past times $t_0$.
Note: it is straightforward to show that $\theta_{\text{DDT}}$ does {\it not} satisfy an ``effective" advection-diffusion equation, even if $s(\vec{x})$ is steady.

This approximation models each individual tracer particle's position by a Gaussian probability distribution with the proper variance (and here, with mean zero, although mean displacements could be included as well).
Dispersion-Diffusion Theory generalizes Batchelor's 1949 theory \cite{B49} for stirring by homogeneous turbulence by retaining the initial position dependence in the effective diffusivity; this matters for inhomogeneous flows such as the steady sinusoidal shear flow.

The exponential decay in the kernel (\ref{rhosol}) suggests that the integral (\ref{ansatz}) may be dominated by the behavior of $\rho$ for small $t-t_0$.  This is the mathematical implementation of the physical statement that the bulk features of the scalar field are determined by the particles most recently injected and depleted by $s$, although this feature is not uniform in the wavenumbers $\vec{k}$ (i.e., in the relevant length-scales).

To evaluate the DDT approximation we define
\begin{equation}\tilde{f}(y)+i\,\tilde{g}(y):=\frac{\sqrt{2}\,k_uS}{2\pi}\sum_{n\in\mathds{Z}}\int_0^\infty dt\int_{0}^{\frac{2\pi}{k_u}}e^{ink_u(y-y_0)-\frac12\,(k_s,nk_u)\, \mathbf{C}(y_0,t)\,(k_s,nk_u)^T}dy_0.
\label{deffandg}
\end{equation}
Then from (\ref{rhosol}), (\ref{ansatz}) and (\ref{Eentries11}-\ref{Eentries22}) we have
\begin{equation}\theta_{\text{DDT}}(x,y)= \tilde{f}(y)\sin(k_sx)+\tilde{g}(y)\cos(k_sx)
\label{apprsol}
\end{equation}
with real functions
\begin{equation}\tilde{f}(y)=\frac{\sqrt{2}\,S}{Uk_s}\sum_{n=-\infty}^{\infty}\cos(nk_uy)I_n, \quad \tilde{g}(y)=\frac{\sqrt{2}\,S}{Uk_s}\sum_{n=-\infty}^{\infty}\sin(nk_uy)I_n
\label{fgexp}
\end{equation}
and the dimensionless integrals
\begin{equation}I_n=\frac{k_uk_sU}{2\pi}\int_0^\infty dt\int_{0}^{\frac{2\pi}{k_u}}\cos(nk_uy_0)e^{-\frac{1}{2}(k_s,nk_u)\, \mathbf{C}(y_0,t)\,(k_s,nk_u)^T}dy_0, \;n\in\mathds{Z}.
\label{defI}
\end{equation}

As an approximation to the exact solution $\theta_\infty$, we use $\theta_{\text{DDT}}$ to compute the multiscale mixing measures and to evaluate its ability to recover the parameter dependences of the efficacies for large P$\acute{\text{e}}$clet number and/or large scale separation.
From (\ref{deffandg}) through (\ref{defI}) and Parseval's Formula, the approximate multi-scale mixing efficacies are
\begin{equation}\label{EsandIs}
\begin{split}
\mathcal{E}_{0}&=\,\sqrt{\frac{\langle\theta_0^2\rangle}{\langle\theta_{\text{DDT}}^2\rangle}}=\Bigg(\frac{\frac{S^2}{\kappa^2k_s^4}}
 {\frac{k_u^2}{4\pi^2}\int_0^{2\pi/k_u}[\tilde{f}^2(y)+\tilde{g}^2(y)]dy}\Bigg)^{\frac{1}{2}}=\frac{r\,\textit{Pe}}{\sqrt{\sum_nI^2_n}}\,,\\
\mathcal{E}_{+1}&=\,\sqrt{\frac{\langle|\nabla\theta_0|^2\rangle}{\langle|\nabla\theta_{\text{DDT}}|^2\rangle}}=\frac{r\,\textit{Pe}}{\sqrt{\sum_n(1+r^2n^2)I^2_n}}\,,\\
\mathcal{E}_{-1}&=\,\sqrt{\frac{\langle|\nabla^{-1}\theta_0|^2\rangle}{\langle|\nabla^{-1}\theta_{\text{DDT}}|^2\rangle}}=\frac{r\,\textit{Pe}}{\sqrt{\sum_n\frac{I^2_n}{1+r^2n^2}}}
\end{split}
\end{equation}
where the sums $\sum_n(\cdot)$ are taken over all integer values of $n$.

In the limit of large scale separation, $r=k_u/k_s\to\infty$, a straightforward change of variables to $z=k_uy_0$ and $\tau=\frac{U^2k_s^2 t}{\kappa k_u^2}$ in (\ref{defI}) using (\ref{Eentries11}-\ref{Eentries22}) suggests
\begin{equation}\sum\nolimits_nI_n^2
\sim I^2_0=\frac{r^2}{\textit{Pe}^2}. \end{equation}
Thus, the intermediate and large scale mixing efficacies approximated by $\theta_{\text{DDT}}$ satisfy
\begin{equation}\mathcal{E}_{0}\sim\mathcal{E}_{-1} \sim\textit{Pe}^2\label{asympht}
\end{equation}
in the limit $r\to\infty$ for large but finite $\textit{Pe}$ in agreement with the homogenization theory prediction $1+\textit{Pe}^2$.

\begin{figure}
  \centering
  \includegraphics[width=8cm,height=7.5cm]{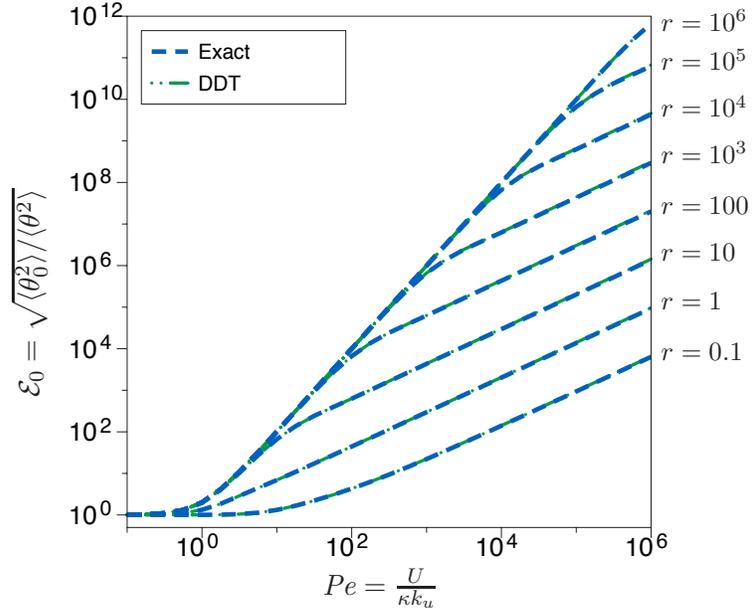}
  \vskip -2.6cm
  \hskip -9cm
  \rotatebox{90}{\makebox[0pt][l]{\large $\mathcal{E}_0=\sqrt{\langle\theta_0^2\rangle/\langle\theta^2\rangle}$}}
  \vskip -5.3cm
  \hskip 7.8cm
  \makebox[0pt][l]{\normalsize $r=10^6$}
  \vskip 0.1cm
  \hskip 7.8cm
  \makebox[0pt][l]{\normalsize $r=10^5$}
  \vskip 0.2cm
  \hskip 7.8cm
  \makebox[0pt][l]{\normalsize $r=10^4$}
  \vskip 0.2cm
  \hskip 7.8cm
  \makebox[0pt][l]{\normalsize $r=10^3$}
  \vskip 0.25cm
  \hskip 7.8cm
  \makebox[0pt][l]{\normalsize $r=100$}
  \vskip 0.2cm
  \hskip 7.8cm
  \makebox[0pt][l]{\normalsize $r=10$}
  \vskip 0.23cm
  \hskip 7.8cm
  \makebox[0pt][l]{\normalsize $r=1$}
  \vskip 0.25cm
  \hskip 7.8cm
  \makebox[0pt][l]{\normalsize $r=0.1$}
  \vskip 2.7cm
  \hskip -1.2cm
  \makebox[0pt][l]{\large $\textit{Pe}=\frac{U}{\kappa k_u}$}
  \caption{\label{e0plotddt}$\mathcal{E}_0$ vs $\textit{Pe}$, for $r=10^m,\;m=-1,0,\cdots,6.$}
\end{figure}

As shown in Figure \ref{e0plotddt}, the dispersion-diffusion approximation to the efficacy (DDT) is visually indistinguishable from the exact value.
While the homogenization theory prediction (HT, dot-dashed line) applies up to $\textit{Pe} \approx r$ and the internal layer asymptotic approximation (ILT, dotted line) is accurate for $\textit{Pe} > \max\{r,1\}$, the dispersion-diffusion approximation is uniformly accurate.

\begin{figure}
  \centering
  \hskip -7cm
 \makebox[0pt][l]{\small $r=10\quad r=100\quad r=10^3\quad r=10^4\quad r=10^5\quad r=10^6$}
 \vskip 0.0cm
 \hskip 0.2cm
  \includegraphics[width=8cm,height=7.5cm]{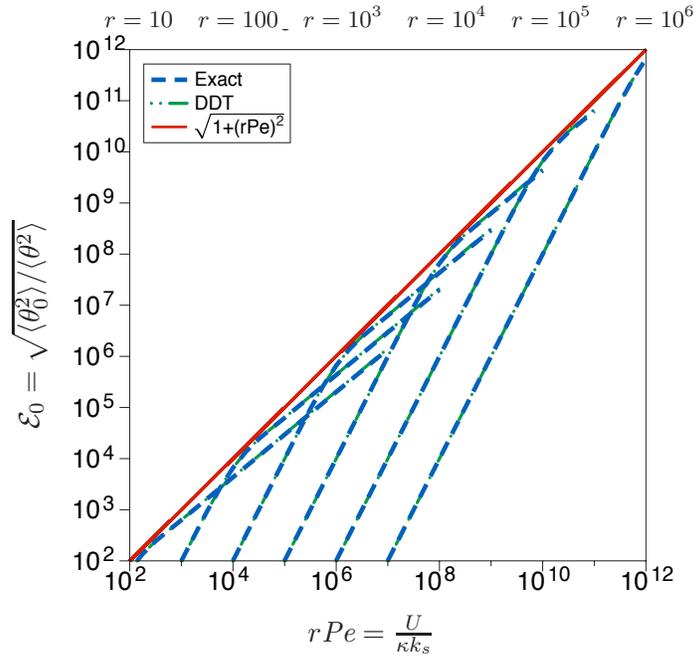}
  \vskip -2.6cm
  \hskip -9cm
  \rotatebox{90}{\makebox[0pt][l]{\large $\mathcal{E}_0=\sqrt{\langle\theta_0^2\rangle/\langle\theta^2\rangle}$}}
  \vskip 2.4cm
  \hskip -1.6cm
  \makebox[0pt][l]{\large $r\textit{Pe}=\frac{U}{\kappa k_s}$}
  \caption{\label{e0plotrpe}$\mathcal{E}_0$ and the rigorous upper bound (solid line) vs $r\textit{Pe}$, for $r=10^m,\;m=-1,0,\cdots,6.$}
\end{figure}

Note that the dispersion-diffusion approximation (along with the exact results, of course) respect the rigorous efficacy bound (\ref{boundrpe}).
The efficacies for a range of $r$ are plotted vs. $r \textit{Pe}$ in Figure \ref{e0plotrpe}.  It is interesting to observe that the simple sine flow nearly saturates the absolute upper bound, which holds for \emph{all} possible stirring flows, when $r \approx \textit{Pe}$.
We expect the dispersion-diffusion approximation to respect the efficacy bound more generally as well.
This is because at high $\textit{Pe}$ we expect the major contribution to $\theta_{\text{DDT}}(\vec{x},t)$ in (\ref{ansatz}) to come from integration times $t_0$ within $l_s/U$ of $t$.
Tracer particle position variance may reasonably be (upper) estimated by
\begin{equation}
\textbf{E}[(X_i(t)-X_i(0))(X_j(t)-X_j(0))] \lesssim (2\kappa t+U^2t^2)
\label{shorttime}
\end{equation}
as $t \to 0$, so for steady sources and sinks each Fourier mode may be estimated
\begin{equation}
|\hat{\theta}_{\text{DDT}}(k)| \gtrsim \frac{|\hat{s}(k)|}{kU}
\label{next}
\end{equation}
as $\textit{Pe} \to \infty$.
This implies that
\begin{equation}
\mathcal{E}_{0}^{DDT} \ \lesssim \ \frac{U l_{source}}{\kappa}
\ = \ \frac{l_{source}}{l_u} \times \frac{U l_u}{\kappa}
\ = \ r \textit{Pe}
\label{more}
\end{equation}
where the distinguished length-scale characterizing a general source-sink distribution is
\begin{equation}
l_{source}^2 \approx \frac{\langle (\Delta^{-1} s)^2 \rangle}
{\langle |\nabla^{-1} s|^2 \rangle},
\label{more2}
\end{equation}
as long as it is non-vanishing.  Should $l_{source}$ vanish, i.e., if the source-sink distribution contains too many small length-scale components, we would expect different (sublinear) high-$\textit{Pe}$ scaling \cite{doering}.

\begin{figure}
  \centering
  \includegraphics[width=8cm,height=7.5cm]{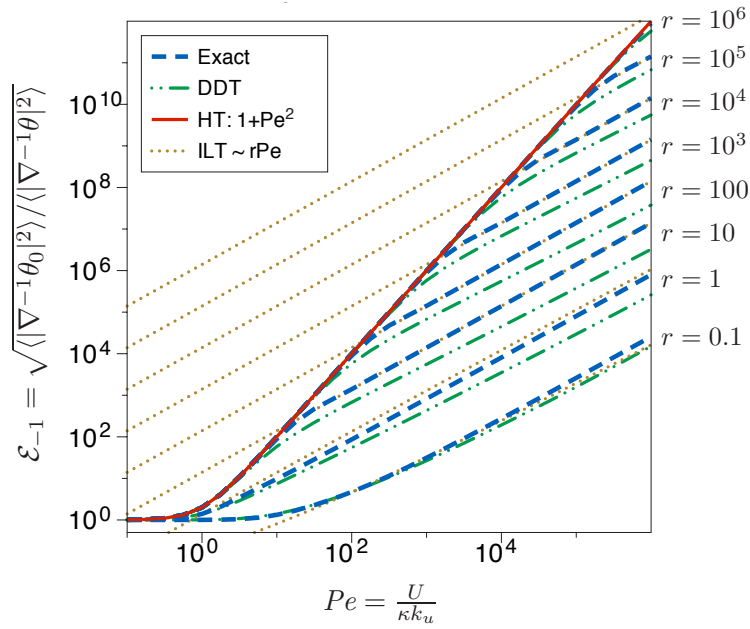}
  \vskip -1.8cm
  \hskip -9cm
  \rotatebox{90}{\makebox[0pt][l]{\large $\mathcal{E}_{-1}=\sqrt{\langle|\nabla^{-1}\theta_0|^2\rangle/\langle|\nabla^{-1}\theta|^2\rangle}$}}
  \vskip -6.2cm
  \hskip 7.8cm
  \makebox[0pt][l]{\normalsize $r=10^6$}
  \vskip 0.1cm
  \hskip 7.8cm
  \makebox[0pt][l]{\normalsize $r=10^5$}
  \vskip 0.15cm
  \hskip 7.8cm
  \makebox[0pt][l]{\normalsize $r=10^4$}
  \vskip 0.16cm
  \hskip 7.8cm
  \makebox[0pt][l]{\normalsize $r=10^3$}
  \vskip 0.16cm
  \hskip 7.8cm
  \makebox[0pt][l]{\normalsize $r=100$}
  \vskip 0.15cm
  \hskip 7.8cm
  \makebox[0pt][l]{\normalsize $r=10$}
  \vskip 0.2cm
  \hskip 7.8cm
  \makebox[0pt][l]{\normalsize $r=1$}
  \vskip 0.35cm
  \hskip 7.8cm
  \makebox[0pt][l]{\normalsize $r=0.1$}
  \vskip 3.1cm
  \hskip -1.2cm
  \makebox[0pt][l]{\large $\textit{Pe}=\frac{U}{\kappa k_u}$}
  \caption{\label{e_1plot}$\mathcal{E}_{-1}$ vs $\textit{Pe}$, for $r=10^m,\;m=-1,0,\cdots,6.$}
\end{figure}

\begin{figure}
  \centering
  \includegraphics[width=8cm,height=7.5cm]{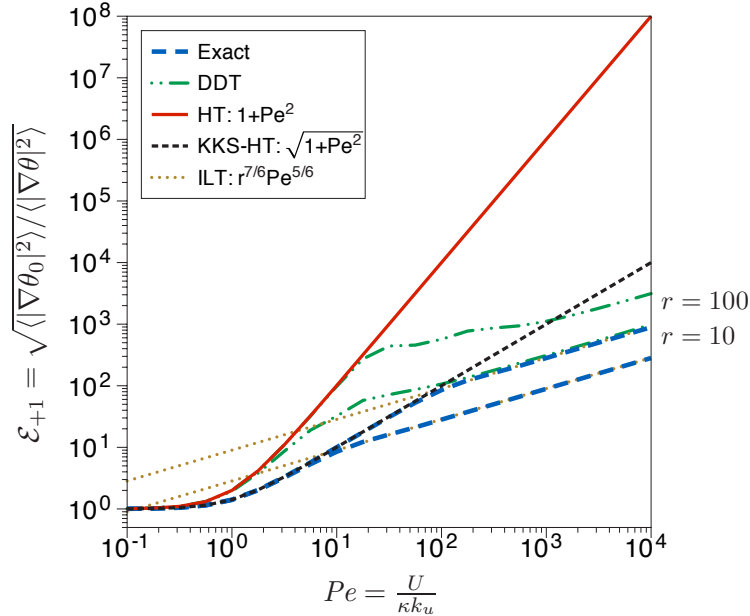}
  \vskip -2cm
  \hskip -9cm
  \rotatebox{90}{\makebox[0pt][l]{\large $\mathcal{E}_{+1}=\sqrt{\langle|\nabla\theta_0|^2\rangle/\langle|\nabla\theta|^2\rangle}$}}
  \vskip -2.2cm
\hskip 7.8cm
\makebox[0pt][l]{\normalsize $r=100$}
\vskip 0.05cm
\hskip 7.8cm
\makebox[0pt][l]{\normalsize $r=10$}
  \vskip 3cm
  \hskip -1.2cm
  \makebox[0pt][l]{\large $\textit{Pe}=\frac{U}{\kappa k_u}$}
  \caption{\label{e1plot}$\mathcal{E}_{1}$ vs $\textit{Pe}$, for $r=10,\,100.$
  For large $r$,
  $I_n$ in (\ref{EsandIs}) is very sensitive to numerical errors in evaluating
  $\mathcal{E}_{+1}$ producing the artificial fluctuations seen in the DDT curves.
  The KKS-HT curve refers to the homogenization theory result for the gradient of
  the scalar field by Keating, Kramer, and Smith \cite{KKS}.  }
\end{figure}

To evaluate the potential for the various theories to capture the mixing efficacies at large and small scales, we plot  the different $\mathcal{E}_{-1}$ and $\mathcal{E}_{+1}$ and their approximations in Figure \ref{e_1plot} and \ref{e1plot}.
For the large-scale mixing efficacy $\mathcal{E}_{-1}$, the homogenization approximation is accurate up to $\textit{Pe} \sim r$ while the dispersion-diffusion approximation appears to capture the correct scaling for any $\textit{Pe}$ modulo a constant prefactor error.
The direct homogenization approximation fails to capture the correct behavior of the small-scale efficiency $\mathcal{E}_{+1}$ for any $\textit{Pe}>1$, which is not unexpected since it explicitly neglects small scale structure in the scalar field.
But a careful homogenization analysis focusing on the gradient of the scalar field \cite{KKS} predicts $\mathcal{E}_{+1} \sim \textit{Pe}$ in the $1 < \textit{Pe} < r$ regime and this {\it does} correctly capture the intermediate behavior.
The DDT approximation, on the other hand, appears to follow the direct (na\"ive) homogenization approximation for $\textit{Pe} \lesssim r$ but then adopts the correct scaling when $\textit{Pe}\gg r$, albeit with a prefactor error.

\begin{figure}
\centering
\includegraphics[width=13cm,height=6.5cm]{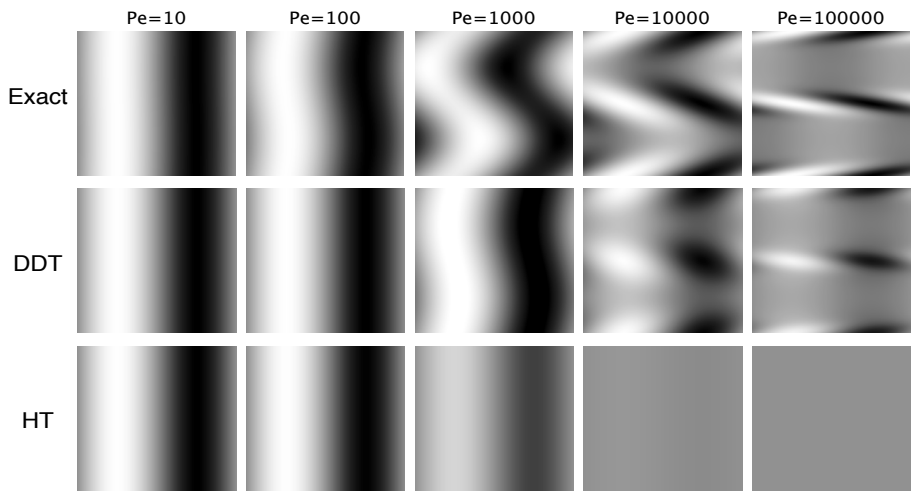}
\caption{\label{fieldplot}Scalar fields for $(x,y)\in[0,2\pi/k_s]\times[0,2\pi/k_u]$ when $r=1000$.}
\end{figure}

To visualize the detailed structures in the scalar field as captured by the various approximations, Figure \ref{fieldplot} compares the exact $\theta_\infty$ and the $\theta_{\text{DDT}}$ and $\theta_{\text{HT}}$ approximations for $\textit{Pe}=10^m$, $m=1,\cdots,5$ for fixed $r=1000$.
The panels in each column (fixed $\textit{Pe}$) are plotted in the same grayscale from $-1$ (black) to $1$ (white) where the scalar fields are normalized by the magnitude of the sup-norm of the corresponding exact solution, $\|\theta_\infty\|_\infty$.
As P$\acute{\text{e}}$clet increases with the intensity of the flow, internal layers develop in the strongly-sheared regions where the speed is relatively small; in the weakly-sheared regions, however, the speed is large and the scalar field is well mixed and has small variance.
It is clear from the figure that although the dispersion-diffusion approximation fails to capture the details of the layers, it does reveal the bulk behavior of the scalar field and more importantly, recovers the correct bulk scalar variance.
The homogenization theory approximation greatly overestimates the effect of the stirring when $\textit{Pe}>r$.

\medskip

\section{Conclusion}
By direct comparison of various mathematical models and measures of mixing we have shown that different definitions of effective diffusion predict distinct large $\textit{Pe}$ scalings of flow-enhanced diffusion.
The discrepancies result from the diverse physical mechanisms motivating the different definitions of the mixing measures.  In particular,

\smallskip

\begin{itemize}
\item In the transient mixing problem ($s\equiv 0$) the long-time behavior of the well-dispersed scalar density is controlled by the advection and diffusion of tracer particles ignoring scalar density structure on length-scales of the stirring.  As a result, the effect of the stirring may be described by a diffusion equation with an effective diffusion tensor enhanced by as much as a factor of $\textit{Pe}^2$.

\smallskip

\item In the presence of sources and sinks, the scalar concentration generally depends on the length-scales of both the flow and the source-sink distribution.  We cannot generally use the long-time, large length-scale dispersion results to approximate the system with an effective diffusion equation: homogenization theory is not applicable without the pristine separation of scales between the stirring and the source-sink distribution.  The enhancement of molecular diffusion by stirring generally depends in a nontrivial way on both $\textit{Pe}$ and the scale separation $r$.

\smallskip

\item Dispersion-Diffusion Theory, motivated by the desire to utilize the essential information in the particle dispersion process in the presence of sources and sinks, reconciles the non-commutative limiting procedures adopted in the literature.  DDT retains the dynamical and inhomogeneous aspects of the effective diffusivity tensor used in homogenization approach and approximates the scalar concentration with an integral similar to the solution of a diffusion equation with sources and sinks.

\smallskip

\item The Dispersion-Diffusion approximation should generally respect the upper bound in (\ref{boundvarred}).  Indeed, at high P\'eclet number the dominant contribution to $\theta_{DDT}$ comes from the most recent times which, due to the gaussian nature of the approximation, leads to variance suppression $\lesssim Pe^1$.

\end{itemize}

\smallskip

We may thus utilize the classical particle dispersion perspective to accurately predict enhanced mixing via scalar variance suppression, even for highly anisotropic and inhomogeneous flows and {\it without} a separation of length-scales.
This is crucial for problems where the sources and sinks possess the same range of scales as the stirring.
We do not, however, yet see how to uniformly reconcile the predictions of the flux-gradient model in (\ref{mastereqfg}).
This model is frequently adopted as the defining framework for turbulent stirring and mixing, but it incorporates infinite scale separation from the start so no such reconciliation may be possible.
This raises the question of the relevance of the flux-gradient model to applications involving statistically steady state mixing in the presence of sources and sinks.

Dispersion-Diffusion Theory may be applied to more general source-sink distributions \cite{doering} and/or more complicated flows like homogeneous and isotropic turbulence where, following Batchelor {\cite{B49}, the dispersive behavior of passive particles is modeled by
\begin{equation}\textbf{E}[ (X_i(t)-X_i(0))(X_j(t)-X_j(0))]\sim (2\kappa t+U^2t^2+\dots)\delta_{ij} \; ,
\label{richrelation}
\end{equation}
at least for displacements within the inertial range.
The term in the covariance $\sim t$ is due to molecular diffusion while the term $\sim t^2$ characterizes the short term drift.
For source-sink distributions with a well-defined spatial scale falling below some ``outer" scale $l_u$ of the turbulence (i.e., $r \lesssim 1$), a calculation very similar to that in (\ref{shorttime}) through (\ref{more2}) with the simple dispersion relation (\ref{richrelation}) yields
\begin{equation}\mathcal{E}_p\sim r\,\textit{Pe},
\label{boundrich}
\end{equation}
for $\textit{Pe} \gg 1 \gtrsim r$, which saturates the scaling of the rigorous upper bounds.  This suggests that homogeneous isotropic turbulence may be a nearly optimal mixer in this sense of steady state scalar variance reduction.
An important aspect of the future research is to test this conjecture with direct numerical simulations and/or experiments for passive scalars that are sustained by steady sources and sinks while being stirred by turbulent flows.


\section*{Acknowledgments}
We gratefully acknowledge stimulating and helpful discussions with Peter R. Kramer, Richard M. McLaughlin, Roberto Camassa, William R. Young and Jean-Luc Thiffeault.  This work was supported in part by NSF awards DMS--0553487, PHY--555324, and PHY--0855335, and also by the Geophysical Fluid Dynamics program at Woods Hole Oceanographic Institution which is supported by NSF and ONR.

\medskip
Received August 2009, revised November 2009.
\medskip
\end{document}